\documentstyle[aps]{revtex}
\def\be{\begin{eqnarray}}
\def\ee{\end{eqnarray}}  
\newcommand{\ep}{\varepsilon}
\newcommand {\bp} {\bbox{p}}
\newcommand {\bq} {\bbox{q}}
\newcommand {\bk} {\bbox{k}}
\newcommand {\bK} {\bbox{K}}
\newcommand {\bv} {\bbox{v}}
\newcommand {\br} {\bbox{r}}
\newcommand {\pp} {\parallel}
\newcommand {\half} {\frac{1}{2}}
\begin{document}
\twocolumn[\hsize\textwidth\columnwidth\hsize\csname %
@twocolumnfalse\endcsname
\draft

\title{Damping of differential rotation in neutron stars}
\author{Armen  Sedrakian}
\address{\it Center for Radiophysics and Space Research
and Department of Astronomy, Cornell University, Ithaca, 
NY 14853}
\date{\today}
\maketitle

\begin{abstract}

\noindent
We derive the transport relaxation times for 
quasiparticle-vortex scattering processes 
via nuclear force, relevant for 
the  damping of differential rotation of superfluids 
in the quantum liquid core of a neutron star. 
 The proton scattering off the neutron vortices  provides 
the dominant resistive force on the vortex lattice at all 
relevant temperatures in the phase where neutrons only are 
in the paired state. If protons are superconducting, 
a small fraction of hyperons and resonances in the normal 
state  would be the dominant source of friction 
on neutron and proton vortex lattices at the core 
temperatures $T\ge 10^{7}$~K.
\end{abstract}
\vspace*{\baselineskip}
]

A broad class of problems related to the nonequilibrium spin 
dynamics of pulsars require the knowledge of the fraction 
of neutron star fluid interiors  which is 
coupled to the crusts on short (unobservable) time scales. 
In practice, one needs  to identify 
the fraction of the moment of inertia of the superfluid in the 
core of the star which is responding to spin perturbations as a rigid body.
A large amount of matter in neutron star models based on the 
modern, moderately soft, equations of state (EoS)   resides at and above 
twice the nuclear matter density. The knowledge 
of the state of the matter above these densities 
is limited, in particular with respect to nucleonic pairing, 
strangeness content, etc.; thus one is led to consider several 
admissible variants of the composition of matter 
when treating the problem of transport. Here we shall adopt 
this type of an approach and examine phases with purely 
nucleonic constituents and those containing 
hypernuclear matter at large densities.

Several {\it electromagnetic} channels of interaction between the 
electrons and the  nucleonic superfluids in the quantum liquid 
cores of neutron stars have been invoked as 
mechanisms of coupling of the superfluid to the normal 
matter \cite{FEIBELMAN,SAULS,ALS,AS,SED}. 
Relativistic electrons couple to the 
anomalous magnetic moment of the neutron quasiparticles localized in the 
cores of vortex lines \cite{FEIBELMAN}. At the densities of interest,
the pairing in neutron matter is driven by the attractive $P$-wave
interaction;  a neutron vortex in a rotating $^3P_2$ superfluid 
acquires an intrinsic magnetization due to the pairing in the $l= 1$ state\cite{SAULS}.
The electron coupling to this magnetization dominates the coupling 
to the neutron magnetic moment  at low temperatures\cite{SAULS}. 
If protons are superconducting even larger magnetization comes about from 
the entrainment effect\cite{ALS,SED}; 
in addition the proton superconductor in the 
mixed state might mediate the coupling between the 
superfluid neutron and electron fluids\cite{SED}.

In this Rapid Communication we examine the 
{\it nuclear interaction} channels of coupling 
between the neutron vortex lines and the bulk quasiparticles; we show that (1)
when protons are normal their scattering off the neutron vortex lines
dominates electromagnetic coupling channels
at all relevant densities and temperatures; and 
(2) if protons are superconducting, the scattering of normal 
(nonsuperfluid) heavy baryons off vortices 
dominates the electromagnetic coupling at the core temperatures $T\ge 10^7$ K.

To start with, consider the quasiclassical Landau-Boltzmann kinetic equation 
for the proton  quasiparticle distribution function
\be\label{1} 
&&\left\{\frac{\partial}{\partial t}
+\frac{\partial\ep_p}{\partial\bp} ~\frac{\partial}{\partial\br}
 - \frac{\partial\ep_p}{\partial \br}~ 
\frac{\partial}{\partial \bp}\right\}~f(\bp, \br, t)\nonumber \\ 
&=&[1-f(\bp, \br, t)]\Sigma^<(\bp, \br, t)
 - f(\bp, \br, t) \Sigma^>(\bp, \br, t), 
\ee
where $\ep_p$ is the proton quasiparticle energy and other variables 
have their usual meaning. The transport vertex  
in the quasiparticle limit is given by
\be\label{2}
\Sigma^<(\bp, \br, t)&=& \frac{n_v^{(n)}}{V} \sum_{\bp', \bq, \bq'}
W(\bp'\bq',\bp\bq)\, \delta(\ep_p+ \ep_q-\ep_{p'}-\ep_{q'})\nonumber \\
&\times& f(\bp',\br,t) \, f(\bq',\br,t)\, 
[1-f(\bq,\br,t)],
\ee
where  $W(\bp'\bq',\bp\bq)$ is the transition probability, 
$\bp$ and $\bq$ denote the proton and neutron momenta, respectively,
$n_v^{(n)}$ is the neutron vortex density per unit area, and $V$ 
is the dimensional normalization. 
The  expression for scattering-out rate, $\Sigma^>(\bp, \br, t)$, follows
from Eq. (\ref{2}) by interchanging the particle and hole occupations.

Microscopically, the neutron vortex lattice is a collection of spatially 
localized normal quasiparticles, which are a superposition of plain waves
along the vortex circulation and two-dimensional bound states 
in the perpendicular 
direction. The continuum of Bardeen-Cooper-Schrieffer (BCS) 
paired state (neutron condensate) feels
the intervening configuration space.
The transition probability for scattering of protons off these localized 
states (vortex cores) is given by
\be\label{3}
W(\bp'\bq',\bp\bq)&=&\frac{2\pi}{\hbar}
\Big\vert G^R(\bp'\bq',\bp\bq;\ep_p+ \ep_q)\Big\vert^2
{\cal S} (\bq,\bq') \nonumber \\ 
&\times&2\pi\hbar \delta(p_{\pp}-p'_{\pp}+q_{\pp}-q'_{\pp}), 
\ee
where the structure factor of the vortex core quasiparticles is 
expressed through the coherence factors $(u,v)$ in the configuration 
space as 
\be\label{4} 
{\cal S}^{1/2}(\bq,\bq') &=& \int \left[u^*_{\bq}(\br_{\perp}) 
u_{\bq'}(\br_{\perp}) - v_{\bq'}(\br_{\perp})
v^*_{\bq}(\br_{\perp}) \right] \nonumber \\
&\times& {\rm exp}\left[i(\bq_{\perp}'-\bq_{\perp})\, \br_{\perp}\right]\, 
d\br_{\perp},
\ee
where $\pp$ and $\perp$ denote components  along  and 
perpendicular to the vector of vortex circulation. 
The expression (\ref{3}) relies on the Brueckner $G$-matrix 
approximation for the transport vertices ($G^R$ is respectively
the retarded component) which involve the resummation of the 
perturbation series in the particle-particle channel.

Considering a situation not far from equilibrium we linearize the
kinetic equation (\ref{1}) with respect to small departure of
proton distribution function. The neutron 
distribution functions are assumed to
be the equilibrium Fermi distributions because the $n$-$n$
relaxation time scales (in the relevant normal state) are much shorter 
than other time scales due to the large number density of 
neutrons. A variational method \cite{FEIBELMAN}
can be used in order to arrive at the relaxation time 
\be\label{5} 
\tau_{pn}^{-1} &=& \frac{n_v^{(n)}}{V} \, \sum_{\bp,\bp',\bq,\bq'}
\frac{1}{2} \left[\psi(\bp)-\psi(\bp')\right]^2
W(\bp'\bq',\bp\bq)\nonumber \\ 
&\times& f(\bp') f(\bq') \left[1-f(\bp)\right] \left[1-f(\bq)\right]\,
{\cal D}^{-1}
\ee 
where 
\be\label{6} 
{\cal D} = \sum_p\psi(\bp)^2\, f(\bp) \, \left[1-f(\bp)\right],
\ee
$\psi(\bp)$ is the trial functon, and 
all the distribution functions are the equilibrium ones. For our 
purposes (macroscopic momentum exchange between the quasiparticle
current and the vortex lattice) 
a good trial function is $\psi(\bp)  = \bp \cdot \bv_L$,
where $\bv_L$ is the vortex velocity. In this case $
{\cal D} = {\nu(p_F) p_F^2 v_L^2}/{3\beta}$,
where $\nu(p_F) = m_p^*p_F/\pi^2\hbar^3$ is the density of states of 
protons at the Fermi surface, $p_F$ is the Fermi momentum, 
$ m_p^*$ is the effective mass of protons,
and $\beta$ is the inverse temperature.
Using the relation 
$f(\omega_1)[1-f(\omega_2)] = g(\omega_1-\omega_2) 
[f(\omega_1)-f(\omega_2)]$,  where $f(\omega)$ and $g(\omega)$ are the 
Fermi and Bose distribution functions, Eq. (\ref{5}) can be rearranged:
\be\label{8}  
\tau_{pn}^{-1} &= &\frac{n_v^{(n)}}{V} \, \sum_{k} 
\frac{1}{2} (\bk\cdot \bv_L)^2 \nonumber \\
&\times&
{\cal D}^{-1}\int_{-\infty}^{\infty}\!\!d\omega \, 
A^{(p)}(\bk,\omega) A^{(n)}(\bk,\omega),
\ee
where the spectral functions of protons and neutrons are 
\be\label{9}  
A^{(p)}(\bk,\omega) &=& \frac{2\pi}{\hbar}\sum_{\bK} \vert G^{R}(\bK+\bk/2, 
\bK-\bk/2;\ep_{\bp_+}+\ep_{\bp_-})\vert^2 \nonumber \\
&\times& \left[ f(\ep_{\bp_+}) - f(\ep_{\bp_-})\right]\delta(\ep_{\bp_+}-\ep_{\bp_-}+\omega),\\
\label{10}
A^{(n)}(\bk,\omega) &=& \sum_{\bq} {\cal S}\left(\bk_{\perp}; 
q_{\pp}+k_{\pp}/2,q_{\pp}-k_{\pp}/2\right)\nonumber \\
&\times&\left[ f(\ep_{\bq_+}) - f(\ep_{\bq_-})\right]\, 
\delta(\ep_{\bq_+}-\ep_{\bq_-}+\omega),
\ee
where $\bk = \bp-\bp'$ and $\bK = (\bp + \bp')/2$ are the 
momentum transfer and 
the center-of-mass momentum, $\bp_{+/-}  = \bK\pm\bk/2$ and $\bq_{+/-}  
= \bq\pm\bk/2$.
The angle intergation in Eq. (\ref{9}) can be carried out using an 
angle averaged on-shell retarded $G$ matrix, which then depends on the 
magnitude of the momentum transfer. To the leading order in 
$\omega/k v_F$ one finds
\be \label{11}
A^{(p)}(\bk,\omega)= \frac{\nu(p_F)}{4}\frac{\omega }{k v_F} \vert\langle 
G^R(k, p_F)\rangle\vert^2\, \theta(2p_F-k),
\ee
where angular brackets denote the angle averaging, 
$v_F$ is the Fermi velocity of protons, and $\theta$ is the Heavyside step 
function. The retarded $G$-matrix can be further 
related to the in-medium $n$-$p$ differential cross section
\be\label{12}
\frac{d\sigma_{pn}}{d\Omega}(k,p_F) = 
\left(\frac{\mu^{*}}{2\pi\hbar^2}\right)^2
\vert\langle 
G^R(k, p_F)\rangle\vert^2,
\ee
where $\mu^*_{pn} = m^*_n m^*_p / (m_n^* + m_p^*)$ is the 
reduced effective mass.

The eigenfunctions of neutron core quasiparticles in the case of $^1S_0$ 
pairing (and in a spinor notation) are \cite{CAROLI,DEGENNES} 
\be\label{13}
\left(\begin{array}{c}
u_{q_{\pp},\mu}(\br_{\perp})\\ v_{q_{\pp}, \mu}(\br_{\perp})\end{array}\right) 
= e^{iq_{\pp}z} \left(e^{i\theta(\mu-\frac{1}{2})}
~e^{i\theta(\mu+\frac{1}{2})}\right) \left(\begin{array}{c}
u'_{\mu}(r)\\ v'_{\mu}(r)\end{array}\right), 
\ee
where $r,\theta, z$ are cylindirical coordinates with the axis of 
symmetry along the vortex circulation, and $\mu$ is the azimuthal 
quantum number, which assumes half-integer positive values. The radial 
functions are 
\be\label{14}
\left(\begin{array}{c}
u'_{\mu}(r)\\ v'_{\mu}(r)\end{array}\right)=2\left(\frac{2}{\pi q_{\perp}r}\right)^{1/2}
e^{-K(r)} \left(\begin{array}{c}
{\rm cos}\left(q_{\perp}r - \frac{\pi\mu}{2} \right) \\ 
{\rm sin}\left( q_{\perp}r - \frac{\pi\mu}{2} \right)\end{array}\right),
\ee
where $q_{\perp} = \sqrt{q^2-q_F^2}$, $q_F$ being the neutron Fermi momentum, 
and 
\be \label{15}
K(r) = \frac{q_F }{\pi q_{\perp} \Delta_{\infty}} \int_0^r\Delta(r') dr'\simeq
\frac{q_F r}{\pi q_{\perp}\xi}\, \left(1 + \frac{\xi e^{-r/\xi}}{r}\right).
\ee
To arrive at the second equality in Eq. (\ref{15})
we assumed a radial dependence of the gap function $\Delta(r) 
= \Delta_{\infty}\, (1-e^{-r/\xi})$ ($\Delta_{\infty}$ 
being the value at $r/\xi \to \infty$), where $\xi$ is the coherence length
of neutron condensate. 
In the low temperature limit, the transitions with a change of the 
azimuthal quantum number can be neglected, and the summation can be 
restricted to the lowest order term $\mu=\mu'=1/2$. The same approximation 
should be used for the eigenvalues of  quasiparticle energies.

Substituting Eqs. (\ref{11})--(\ref{13}) in Eq. (\ref{4}) and taking 
into account the normalization of states, we find
\be \label{16}
{\cal S}_{\half ,\half }^{1/2} (\bk_{\perp}) 
&=& \int\! dy \,  J_0(x,y) 
\,{\rm cos}\left(2 x y -\pi\mu\right)e^{-2K(x,y)} \nonumber \\
&\times&\left\{\int dy   \, 
e^{-2K(x, y)}\right\}^{-1},
\ee
where $J_0$ is the Bessel function, $x = k_{\perp}\xi$ and $y  = r/\xi$.
For a fixed value of $q_F\xi$, the parametrical dependence
of ${\cal S}$ on $x$ can be well fitted by a Lorenzian
after calculating the integrals numerically. For a typical 
value $q_F\xi = 10$, e.g., 
 we find
\be \label{17}
{\cal S}(x) = \frac{a\,\gamma}{(x - x_0)^2 + \gamma^2},
\ee
where the amplitude is $a = 0.11$, the maximum point $x_0 = 1.1$ and the 
width $\gamma = 0.4$. The neutron quasiparticle spectrum is given by
\be \label{18}
\ep_{\mu}(q) &=& \frac{\mu\hbar}{\sqrt{q_F^2-q^2}}\int_0^{\infty}\! dr
\frac{\Delta(r)}{r}e^{-2K(r)} \nonumber 
\\
&\times & \left\{\int_0^{\infty}dr e^{-2K(r)}\right\}^{-1}
 \simeq\ep_{\mu}^0\left( 1+\frac{q^2}{2q_F^2}\right),
\ee
where $\ep_{\mu}^0 = \mu\pi\Delta^2_{\infty}/2\ep_{Fn}$ with $\ep_{Fn}$
being the Fermi energy of neutrons. The second equality is obtained by keeping
the next-to-leading terms in small quantity $q/q_F$ and approximating the integral
as $\Delta_{\infty}/\xi$. With the approximations above 
the spectral function for neutrons in the low temperature limit ($e^{-\beta\omega}\ll 1$) is \cite{FEIBELMAN} 
\be \label{19}
A^{(n)}(\bk, \omega)&=& \frac{q_F^2}{\pi\hbar\ep_{1/2}^0\, k_z}
{\rm exp}\left[-\beta\ep_{1/2}^0\left(1+\frac{\omega^2q_F^2}
{2\ep_{1/2}^{0\, 2}k_z^2 } + \frac{k_z^2}{8q_F^2} \right)\right]
\nonumber \\
&\times&{\rm sinh}\left(\frac{\beta\omega}{2}\right) 
{\cal S}_{\half,\half}(k_{\perp}) .
\ee
We further substitute the spectral 
functions in Eq. (\ref{9}) and perform the $\omega$ integration. 
Further progress can be made if we assume that the scattering amplitude is independent of $k_z$ and carry out the $k_z$ integration. We find
\be \label{20}
\tau_{pn}^{-1} &=& 6 e^{1/2\pi}
\,K_0\left(\frac{1}{2\pi}\right)\, n_v^{(n)}
\left(\frac{m_n^*}{\mu^*_{pn}}\right)
\left(\frac{\ep_{Fn}}{\ep_{Fp}}\right)^2
\nonumber \\
&\times&
\frac{1}{\beta\ep^0_{1/2}}e^{-\beta\ep^0_{1/2}}\frac{\hbar}{m_p^*\xi^2}\,
\frac{d\sigma_{pn}^{*}}{d\Omega} ,
\ee
where 
\be \label{21}
\frac{d\sigma_{pn}^{*}}{d\Omega}
 = \int_{0}^{x_1}\!\! dx\, x^2\, {\cal S}(x)
\frac{d\sigma_{pn}}{d\Omega}(x,p_F) ,
\ee
with  $x_1= 2p_F\xi/\hbar$; here $K_0$ is the modified Bessel function. 
The temperature dependence of the relaxation time is due to 
protons contributing the $\beta^{-1}$ (thermal smearing of the 
Fermi surface) and vortex core quasiparticles contributing the 
$\exp (-\beta \ep_{1/2}^0)$  (the probability 
of exciting a neutron quasiparticle inside the core of a 
neutron vortex). 
The $n$-$p$ scattering cross section is modified in the 
medium,  first,  due to the suppression of the 
intermediate state two-particle 
propagation and, second, due to the modification of the density of states.
We incorporate the medium modifications of the cross section 
using the code of Ref. \cite{SRA} with the separable form of the Paris 
potential and a modified Pauli-blocking operator which keeps
only the intermediate occupation numbers of neutrons and 
sets  $f(\bp) = 0$ (for a discussion of the related issues
see also \cite{SCHULZE}).

Table I shows the density and temperature dependence of the 
relaxation times  $\tau_{en}$ (Ref. \cite{FEIBELMAN}), 
$\tau_{eM}$ (Ref. \cite{SAULS}), and $\tau_{pn}$ (present work).
 The microscopic parameters are from Ref. \cite{BALDO} 
(the proton $^1S_0$ gap, according to  \cite{BALDO}, 
vanishes at $n = 0.4$ fm$^{-3}$, while $^3P_2$ neutron gap persists through 
the whole density region of interest; recent calculations 
of Ref. \cite{ELG}, which  include the tensor coupling,
indicate that superfluid neutron--normal proton mixture exists in the 
density region $0.43 \le n \le 0.45$ fm$^{-3}$  while matter is normal 
at larger densities). 

The ratio of electron--neutron-vortex to proton--neutron-vortex  
relaxation times $\tau_{en}/\tau_{pn}$ is of the order $10^{6}$ and 
mainly reflects the difference in the spin--magnetic-moment and nuclear 
interaction matrix elements. As the temperature decreases, both time scales 
increase and for $\beta\gg \ep_{Fn}/\pi\Delta^2$ tend to infinity 
due to  the exponential cutoff of the scattering rates. Note that 
the relaxation time for electron scattering off
the $P$-wave vortex magnetization, $\tau_{eM}$,
is temperature independent; the crossover 
temperature corresponding to $\tau_{eM}\ge \tau_{pn}$ is 
of the order of $3\times 10^{6}$ K.

Consider now neutron star matter which, in addition 
to the neutron-proton-electron ($npe$) mixture,  
contains also heavy baryons in the
normal state. Among these the lowest threshold densities have the  
$\Sigma^-$ and $\Lambda$ hyperons and the $\Delta^{-}$ resonance. 
The threshold densities in the case
of noninteracting Fermi gases and relativistic mean-field models are 
low, a few times nuclear saturation density 
($n_{\rm sat}=0.17$ fm$^{-3}$)\cite{GLEN,GLEN2}. 
Correlations tend to shift the 
threshold densities to higher values in general. 
\begin{table}
\begin{center}
\caption{ 
The relaxation times  for the  spin period 
of the Vela pulsar $P = 0.089$ s ($\tau$ scale $\propto P$) and 
for temperatures $\beta^{-1} \simeq 10^8,~10^7,$ and 
$~3\times 10^{6}$ K (in descending order). $x_p$ is the proton fraction.
}

\begin{tabular}{cccccccc}
$n$ & $x_p$ & $\Delta$&  $\sigma_{pn}^*$ &$\tau_{pn}$
 &$\tau_{en}$ & $\tau_{eM}$ \\
(fm$^{-3})$&  &  (MeV)&   [fm$^2$]& & &  \\
\\
\hline 
0.4 & 0.060 & 0.7 &  87.2 
 &$\begin{array}{c}0.001^{\rm s}\\ 0.004^{\rm s}\\ 3.670^{\rm s}\end{array}
$                                    & $
\begin{array}{c} 0.027^{\rm d} \\0.108^{\rm d}\\ 11.190^{\rm d}\end{array}
$ & 87.175$^{\rm d}$ \\
\hline
0.6 & 0.070 & 1.3 &  51.7 & $
\begin{array}{c}0.006^{\rm s}\\ 0.015^{\rm s} \\ 41.56^{\rm d}\end{array}
$                    & $
\begin{array}{c}4.516^{\rm d}\\ 12.426^{\rm d}\\ 18947.12^{\rm yr} \end{array}
$  & 51.699$^{\rm d}$  \\
\hline
0.8 & 0.084 & 1.3 &  57.7  & $
\begin{array}{c}0.011^{\rm s}\\ 0.022^{\rm s}\\ 111.71^{\rm d}\end{array}
$                   & $
\begin{array}{c}9.791^{\rm d}\\19.175^{\rm d}\\ 36279.45^{\rm yr}\end{array}
$ & 57.865$^{\rm d}$ \\
\hline
1.0 & 0.090 & 1.2 &  65.1 & $
\begin{array}{c}0.015^{\rm s}\\ 0.024^{\rm s}\\ 23.99^{\rm d}\end{array}
$                   & $
\begin{array}{c}6.876^{\rm d}\\10.811^{\rm d}\\ 397320^{\rm yr}\end{array}
$ & 65.098$^{\rm d}$ 
\end{tabular}
\end{center}
\end{table} 
While the uncertainties in 
the hypernuclear matter calculations are still 
considerable, we shall, for the 
purpose of order of magnitude estimate, make  several simplifying 
assumptions below.
In the high density  $n^{(s)}peY$ phase, where  superscript $s$
stands for the superfluid state and  $Y\equiv\Sigma^-, \, \Lambda\, , 
\Delta^{-}$,  the $Y$-$n$ channel due to the heavy baryons 
would contribute additively  to the scattering rate in the $p$-$n$ channel;  
in this case no new qualitative effects are expected.

The situation is different in the $n^{(s)}p^{(s)}eY$ phase, where in
addition to the neutron condensate, the proton superconductor 
is in the mixed state.   
The relaxation time for the dominant electromagnetic coupling
via scattering of electrons by the magnetic flux 
of a vortex is \cite{ALS,AS,SED}
\be 
\tau_{e\Phi}^{-1} = \frac{3\pi^3}{64}n_v^{(n/p)}\frac{ c\lambda}{(k_{eF} \lambda)^2},
\ee
where $k_{eF}$ is the electron Fermi wave vector 
and $\lambda$ is the penetration
depth of the proton superconductor, $n_v^{(n/p)}$ 
is the density of either neutron or
proton vortex lines. To compare this time with 
the case of nuclear scattering of heavy baryons by neutron 
and proton vortex core quasiparticles we assume in Eq. (\ref{20}) 
(with obvious modifications) $\sigma_{Yn} = 40 $ mb
and replace the effective masses of heavy baryons by their bare masses
$m_{\Lambda}= 1116$, $m_{\Sigma^{-}}=1197$, and  $m_{\Delta^-}=1232$ MeV.
Once the threshold density is achieved, the hyperon population grows 
steeply to about $10\%$ and stays further almost constant\cite{GLEN,GLEN2}.
Table II shows the net relaxation time 
 [Eq. (\ref{20})] due to $ \Lambda\, , \Sigma^-\, ,
\Delta^{-}$ scattering off neutron or proton quasiparticles localized in the
vortex cores   at the threshold density 
$n = 0.4$ fm$^{-3}$, and for a well developed hyperon core at  $n = 0.5$ fm$^{-3}$. The relative abundances for heavy baryons are assumed equal and
fixed at $1\%$ and $10\%$ for each case respectively. 
The values of $\tau_{Yp}$ are larger than that of $\tau_{Yn}$ at $n = 0.4$ fm$^{-3}$
mainly due to the large ratio of neutron to proton Fermi energies, i.e., 
more abundant neutrons are more effective in scattering of 
hyperons. 
\begin{table}
\begin{center}
\caption{Relaxation times in hypernuclear matter; 
conventions are the same as in Table I.
$x_Y$ is the concentration of each species of heavy baryons 
($\Sigma^- ,\Lambda  ,$ 
and $\Delta^{-}$); $\tau_{Yn}$ and $\tau_{Yp}$ are the 
relaxation times for scattering of all species of heavy baryon off 
neutron and proton vortex core quasiparticles, respectively. The electron--vortex-flux
relaxation time, $\tau_{e\Phi}$, is given per single flux quantum $\phi_0= \pi\hbar e/c$.} 
\begin{tabular}{ccccccccc}
$n$ & $x_p$ & $x_Y$ & $\Delta_p$ & $\Delta_n$ 
&$\tau_{Yp}$ &$\tau_{Yn}$ &$\tau_{e\Phi}$  \\
(fm$^{-3})$&  & & (MeV)&  (MeV) &  & \\
\\
\hline 
0.4 & 0.05 & 0.01 & 0.4 & 0.7 &$\begin{array}{c}0.036^{\rm s}\\ 
70.34^{\rm s}\\292.4^{\rm d}\end{array}$ 
& $\begin{array}{c}0.0002^{\rm s} \\0.024^{\rm s}
\\ 21.18^{\rm s}\end{array}$ 
& 0.167$^{\rm s}$ \\
\hline 
0.5 & 0.1 & 0.1 & 0.1 & 1.0 &
$\begin{array}{c}0.027^{\rm s}\\ 
0.32^{\rm s}\\1.43^{\rm s}\end{array}$ 
& 
$\begin{array}{c}0.0002^{\rm s}\\ 15.15^{\rm s}
\\ 45.94^{\rm d}\end{array}$ 
& 0.195$^{\rm s}$ \\
\end{tabular}
\end{center}
\end{table}
At $n = 0.5$ fm$^{-3}$ the trend is 
opposite mainly due to an order of magnitude difference in the pairing gaps. 
One should note, however, that for 
typical pulsar magnetic fields, $B\sim 10^{12} - 10^{13}$ G,
the number of proton vortices per neutron vortex 
is very large,  $n_v^{(p)}/n_v^{(n)}\sim 10^{12}$--$10^{13}$ and  $\tau_{Yp}$
is the  main quantity of interest. 
For core temperatures  $T\sim 10^8$ K, typical 
for not extremely old neutron stars, 
the hyperon scattering appears to be the dominant 
source of friction. It could 
be also important in the well developed hyperon cores at $T\sim 10^7$ K, 
below the densities at which the proton 
superconductivity is quenched because of 
the high  proton fraction.

In summary, our calculations above show that 
scattering of baryons via nuclear force 
off vortex core quasiparticles in neutron star's superfluid interiors 
may be dominant in a number of cases. In the so-called $n^{(s)}pe$ (or $n^{(s)}peY$) phase, 
where only the neutrons form a condensate, the baryon (including hyperons
and resonances) scattering off neutron vortex core quasiparticles
dominates other channels driven by the electromagnetic 
interactions \cite{FEIBELMAN,SAULS}.
In the  $n^{(s)}p^{(s)}e$ phase, where the type-II proton superconductor 
is in the mixed state, a trace of heavy baryons 
(such as $\Sigma^- ,\Lambda  ,$ 
and $\Delta^{-}$) in the normal state
would produce a resistive force on vortices which will dominate 
at  temperatures $\sim 10^8$ K and will give 
the way to electron flux scattering at lower 
temperatures $\le 10^{7}$ K \cite{ALS,AS,SED}. 
Which of the channels dominates at the intermediate 
temperatures $\sim 10^{7}$ K, depends on the 
details of the microphysical input, such as
the pairing gaps and baryon abundances.
For practical applications, our main 
result, Eq. (\ref{20}), can be cast in a compact form $(\hbar = c = 1)$
\be 
\tau_{pn} = 1.5 
\frac{\mu^*_{pn}}{m_p^*}\left(\frac{k_{Fp}}{k_{Fn}}\right)^4
            \frac{\beta}{\sigma^*_{pn}} \,
{\rm exp}\left(0.038\, \frac{m^*_n}{m_n} 
\frac{ \beta\Delta_n^2}{k_{Fn}^2} \right) P,
\ee
where the relations $n_v^{(n)} = (4m_n/\hbar)  P^{-1}$, where $P$ is 
the spin period,  and $\sigma^*_{pn}/4\pi  = d\sigma^*_{pn}/d\Omega$ 
have been used. 

As an illustrative example 
consider first a neutron star model of canonical mass 1.4 $M_{\odot}$ 
based on the EoS of Ref.\cite{WIRINGA}.
The $n^{(s)}pe$-phase in this model occupies 
the radial range $0\le R\le 8$ km 
for the gap profiles of Ref.\cite{BALDO} and $7 \le R\le 7.7 $ km 
for the profiles of Ref. \cite{ELG} (star radius is 10.13 km).
For the maximum mass, $M =1.65 M_{\odot}$,  star model of Ref. 
\cite{GLEN2} (star radius is 11.4 km),
the $n^{(s)}p^{(s)}eY$-phase occupies the shell $7\le R\le 9.5$ while at smaller
radii the $^1S_0$ proton pairing vanishes  ($n^{(s)}peY$-phase). 
These numbers  
by no means provide a consistent overall picture of the distribution of 
different superfluid phases in a neutron star, for the results 
on the  pairing gaps  and  EoS stem from different 
(BCS-BHF \cite{BALDO,ELG} vs 
respectively variational \cite{WIRINGA} and mean-field \cite{GLEN})
approaches to the nuclear many-body problem. Nevertheless, 
our observation of the dominance of nuclear interaction channels 
when any kind of baryons is in the normal state is robust, i.e., independent 
of the details of the microphysical input. Its {\it implementation},
however, as given by the example above, is connected with the  uncertainties
due to limited knowledge of the interactions and pairing in 
the high density nuclear matter as well as the heavy baryon 
concentrations and their density thresholds.

This research has been supported by the Max Kade Foundation
(New York, NY).

\end{document}